\documentclass[sigconf]{acmart}

\usepackage{enumitem}
\usepackage{listings}
\usepackage{balance}

\newcommand{\ignore}[1]{}

\setcopyright{none}
\renewcommand\footnotetextcopyrightpermission[1]{}

\begin{document}

\title{Pyserini: An Easy-to-Use Python Toolkit to Support Replicable IR Research with Sparse and Dense Representations}

\author{Jimmy Lin, Xueguang Ma, Sheng-Chieh Lin,\\
Jheng-Hong Yang, Ronak Pradeep, and Rodrigo Nogueira}

\affiliation{\vspace{1ex}
David R.\ Cheriton School of Computer Science\\
University of Waterloo\\
}

\begin{abstract}
Pyserini is an easy-to-use Python toolkit that supports replicable IR research by providing effective first-stage retrieval in a multi-stage ranking architecture.
Our toolkit is self-contained as a standard Python package and comes with queries, relevance judgments, pre-built indexes, and evaluation scripts for many commonly used IR test collections.
We aim to support, out of the box, the entire research lifecycle of efforts aimed at improving ranking with modern neural approaches.
In particular, Pyserini supports sparse retrieval (e.g., BM25 scoring using bag-of-words representations), dense retrieval (e.g., nearest-neighbor search on transformer-encoded representations), as well as hybrid retrieval that integrates both approaches.
This paper provides an overview of toolkit features and presents empirical results that illustrate its effectiveness on two popular ranking tasks.
We also describe how our group has built a culture of replicability through shared norms and tools that enable rigorous automated testing.
\end{abstract}

\renewcommand{\shortauthors}{}
\pagestyle{empty}

\maketitle

\section{Introduction}

The advent of pretrained transformers has led to many exciting recent developments in information retrieval~\cite{Lin_etal_arXiv2020_ptr4tr}.
In our view, the two most important research directions are transformer-based re\-ranking models and learned dense representations for ranking.
Despite many exciting opportunities and rapid research progress, the need for easy-to-use, replicable baselines has remained a constant.
In particular, the importance of stable first-stage retrieval within a multi-stage ranking architecture has become even more important, as it provides the foundation for increasingly-complex modern approaches that leverage hybrid techniques.

We present Pyserini, our Python IR toolkit designed to serve this role:\ it aims to provide a solid foundation to help researchers pursue work on modern neural approaches to information retrieval.
The toolkit is specifically designed to support the complete ``research lifecycle'' of systems-oriented inquiries aimed at building better ranking models, where ``better'' can mean more effective, more efficient, or some tradeoff thereof.
This typically involves working with one or more standard test collections to design ranking models as part of an end-to-end architecture, iteratively improving components and evaluating the impact of those changes.
In this context, our toolkit provides the following key features:

\begin{itemize}[leftmargin=*]

\item Pyserini is completely self-contained as a Python package, available via \texttt{pip install}.
The package comes with queries, collections, and qrels for standard IR test collections, as well as pre-built indexes and evaluation scripts.
In short, batteries are included.
Pyserini supports, out of the box, the entire research lifecycle of efforts aimed at improving ranking models.

\item Pyserini can be used as a standalone module to generate batch retrieval runs or be integrated as a library into an application designed to support interactive retrieval.

\item Pyserini supports sparse retrieval (e.g., BM25 scoring using bag-of-words representations), dense retrieval (e.g., nearest-neighbor search on transformer-encoded representations), as well hybrid retrieval that integrates both approaches.

\item Pyserini provides access to data structures and system internals to support advanced users.
This includes access to postings, document vectors, and raw term statistics that allow our toolkit to support use cases that we had not anticipated.

\end{itemize}

\noindent Pyserini began as the Python interface to Anserini~\cite{Yang_etal_SIGIR2017,Yang_etal_JDIQ2018}, which our group has been developing for several years, with its roots in a community-wide replicability exercise dating back to 2015~\cite{Lin_etal_ECIR2016}.
Anserini builds on the open-source Lucene search library and was motivated by the desire to better align academic research with the practice of building real-world search applications; see, for example,~\citet{Grand_etal_ECIR2020}.
More recently, we recognized that Anserini's reliance on the Java Virtual Machine (due to Lucene), greatly limited its reach~\cite{Yilmaz_etal_EMNLP2019demo,Yilmaz_etal_SIGIR2020}, as Python has emerged as the language of choice for both data scientists and researchers.
This is particularly the case for work on deep learning today, since the major toolkits (PyTorch~\cite{paszke2019pytorch} and Tensorflow~\cite{abadi2016tensorflow}) have both adopted Python as their front-end language.
Thus, Pyserini aims to be a ``feature-complete'' Python interface to Anserini.

Sparse retrieval support in Pyserini comes entirely from Lucene (via Anserini).
To support dense and hybrid retrieval, Pyserini integrates Facebook's FAISS library for efficient similarity search over dense vectors~\cite{FAISS}, which in turns integrates the HNSW library~\cite{HNSW} to support low-latency querying.
Thus, Pyserini provides a superset of features in Anserini; dense and hybrid retrieval is entirely missing from the latter.

This paper is organized in the following manner:\ 
After a preamble on our design philosophy, we begin with a tour of Pyserini, highlighting its main features and providing the reader with a sense of how it might be used in a number of common scenarios.
This is followed by a presentation of empirical results illustrating the use of Pyserini to provide first-stage retrieval in two popular ranking tasks today.
Before concluding with future plans, we discuss how our group has internalized replicability as a shared norm through social processes supported by technical infrastructure.

\section{Design Philosophy}

The design of Pyserini emphasizes ease of use and replicability.
Larry Wall, the creator of the Perl programming language, once remarked that ``easy things should be easy, and hard things should be possible.''
While aspects of the lifecycle for systems-oriented IR research are not difficult per se, there are many details that need to be managed:\ downloading the right version of a corpus, building indexes with the appropriate settings (tokenization, stopwords, etc.), downloading queries and relevance judgments (deciding between available ``variants''), manipulating runs into the correct output format for the evaluation script, selecting the right metrics to obtain meaningful results, etc.
The list goes on.
These myriad details often trip up new researchers who are just learning systems-oriented IR evaluation methodology (motivating work such as~\citet{Yilmaz_etal_SIGIR2020}), and occasionally subtle issues confuse experienced researchers as well.\footnote{As a concrete example, TREC-COVID has (at least) 12 different sets of qrels. All of them are useful for answering different research questions. Which one do you use?} 
The explicit goal of Pyserini is to make these ``easy things'' easy, supporting common tasks and reducing the possibility of confusion as much as possible.

At the other end of the spectrum, ``hard things should be possible''.
In our context, this means that Pyserini provides access to data structures and system internals to support researchers who may use our toolkit in ways we had not anticipated.
For sparse retrieval, the Lucene search library that underlies Anserini provides interfaces to control various aspects of indexing and retrieval, and Pyserini exposes a subset of features that we anticipate will be useful for IR researchers.
These include, for example, traversing postings lists to access raw term statistics, manipulating document vectors to reconstruct term weights, and fine-grained control over document processing (tokenization, stemming, stopword removal, etc.).
Pyserini aims to sufficiently expose Lucene internals to make ``hard things'' possible.

Finally, the most common use case of Pyserini as first-stage retrieval in a multi-stage ranking architecture means that replicability is of utmost concern, since it is literally the foundation that complex reranking pipelines are built on.
In our view, replicability can be divided into technical and social aspects:\ an example of the former is an internal end-to-end regression framework that automatically validates experimental results.
The latter includes a commitment to ``eat our own dog food'' and the adoption of shared norms.
We defer more detailed discussions of replicability to Section~\ref{section:replicability}.

\section{Pyserini Tour}

Pyserini is packaged as a Python module available on the Python Package Index.
Thus, the toolkit can be installed via \texttt{pip}, as follows:

\smallskip
\begin{quote}
\begin{verbatim}
$ pip install pyserini==0.11.0.0
\end{verbatim}
\end{quote}

\smallskip
\noindent In this paper, we are explicitly using v0.11.0.0.
The code for the toolkit itself is available on GitHub at \texttt{pyserini.io}; for users who may be interested in contributing to Pyserini, we recommend a ``development'' installation, i.e., cloning the source repository itself.
However, for researchers interested only in {\it using} Pyserini, the module installed via \texttt{pip} suffices.

In this section, we will mostly use the MS MARCO passage ranking dataset~\cite{MS_MARCO_v3} as our running example.
The dataset has many features that make it ideal for highlighting various aspects of our toolkit:\ the corpus, queries, and relevance judgments are all freely downloadable; the corpus is manageable in size and thus experiments require only modest compute resources (and time); the task is popular and thus well-studied by many researchers.

\subsection{Interactive Retrieval}

In Figure~\ref{fig:sparse-retrieval-example}, we begin with a simple example of using Pyserini to perform bag-of-words ranking with BM25 (the default ranking model) on the MS MARCO passage corpus (comprising 8.8M passages).
To establish a parallel with ``dense retrieval'' techniques using learned transformer-based representations (see below), we refer to this as ``sparse retrieval'', although this is not common parlance in the IR community at present.

\begin{figure}[t]
\small
\begin{tabular}{c}
\begin{lstlisting}
from pyserini.search import SimpleSearcher

searcher = SimpleSearcher.from_prebuilt_index('msmarco-passage')
hits = searcher.search('what is a lobster roll?', 10)

for i in range(0, 10):
    print(f'{i+1:2} {hits[i].docid:7} {hits[i].score:.5f}')

\end{lstlisting}
\end{tabular}
\vspace{-0.2cm}
\caption{Simple example of interactive sparse retrieval (i.e., bag-of-word BM25 ranking).}
\label{fig:sparse-retrieval-example}
\vspace{-0.25cm}
\end{figure}

The \texttt{SimpleSearcher} class provides a single point of entry for sparse retrieval functionality.
In (L3), we initialize the searcher with a pre-built index.
For many commonly used collections where there are no data distribution restrictions, we have built indexes that can be directly downloaded from our project servers.
For researchers who simply want an ``out-of-the-box'' keyword retrieval baseline, this provides a simple starting point.
Specifically, the researcher does not need to download the collection and build the index from scratch.
In this case, the complete index, which includes a copy of all the texts, is a modest 2.6GB.

Using an instance of \texttt{SimpleSearcher}, we issue a query to retrieve the top 10 hits (L4), the results of which are stored in the array \texttt{hits}.
Naturally, there are methods to control ranking behavior, such as setting BM25 parameters and enabling the use of pseudo-relevance feedback, but for space considerations these options are not shown here.
In (L6--7), we iterate through the results and print out rank, docid, and score.
If desired, the actual text can be fetched from the index (e.g., to feed a downstream reranker).

Figure~\ref{fig:dense-retrieval-example} shows an example of interactive retrieval using dense learned representations.
Here, we are using TCT-ColBERT~\cite{Lin_etal_arXiv2020_DenseRanking}, a model our group has constructed from ColBERT~\cite{Khattab_Zaharia_SIGIR2020} using knowledge distillation.
As with sparse retrieval, we provide pre-built indexes that can be directly downloaded from our project servers.
The \texttt{SimpleDenseSearcher} class serves as the entry point to near\-est-neighbor search functionality that provides top $k$ retrieval on dense vectors.
Here, we are taking advantage of HNSW~\cite{HNSW}, which has been integrated into FAISS~\cite{FAISS} to enable low latency interactive querying (L6).

\begin{figure}[t]
\begin{tabular}{c}
\begin{lstlisting}
from pyserini.dsearch import SimpleDenseSearcher, \
                             TCTColBERTQueryEncoder

encoder = TCTColBERTQueryEncoder('castorini/tct_colbert-msmarco')
searcher = SimpleDenseSearcher.from_prebuilt_index(
    'msmarco-passage-tct_colbert-hnsw',
    encoder
)
hits = searcher.search('what is a lobster roll')
\end{lstlisting}
\end{tabular}
\vspace{-0.2cm}
\caption{Simple example of interactive dense retrieval (i.e., approximate nearest-neighbor search on dense learned representations).}
\label{fig:dense-retrieval-example}
\vspace{-0.4cm}
\end{figure}

The final component needed for dense retrieval is a query encoder that converts user queries into the same representational space as the documents.
We initialize the query encoder in (L4), which is passed into the method that constructs the searcher.
The encoder itself is a lightweight wrapper around the Transformers library by Huggingface~\cite{wolf-etal-2020-transformers}.
Retrieval is performed in the same manner (L9), and we can manipulate the returned \texttt{hits} array in a manner similar to sparse retrieval (Figure~\ref{fig:sparse-retrieval-example}).
At present, we support the  TCT-ColBERT model~\cite{Lin_etal_arXiv2020_DenseRanking} as well as DPR~\cite{karpukhin-etal-2020-dense}.
Note that our goal here is to provide retrieval capabilities based on existing models; quite explicitly, representational learning lies outside the scope of our toolkit (see additional discussion in Section~\ref{section:future}).

\begin{figure}[t]
\begin{tabular}{c}
\begin{lstlisting}
from pyserini.search import SimpleSearcher
from pyserini.dsearch import SimpleDenseSearcher, \
                             TCTColBERTQueryEncoder
from pyserini.hsearch import HybridSearcher

ssearcher = SimpleSearcher.from_prebuilt_index('msmarco-passage')
encoder = TCTColBERTQueryEncoder('castorini/tct_colbert-msmarco')
dsearcher = SimpleDenseSearcher.from_prebuilt_index(
    'msmarco-passage-tct_colbert-hnsw',
    encoder
)
hsearcher = HybridSearcher(dsearcher, ssearcher)
hits = hsearcher.search('what is a lobster roll', 10)
\end{lstlisting}
\end{tabular}
\vspace{-0.2cm}
\caption{Simple example of interactive search with hybrid sparse--dense retrieval.}
\label{fig:hybrid-retrieval-example}
\vspace{-0.5cm}
\end{figure}

Of course, the next step is to combine sparse and dense retrieval, which is shown in Figure~\ref{fig:hybrid-retrieval-example}.
Our \texttt{HybridSearcher} takes as its constructor the sparse retriever and the dense retriever and performs weighted interpolation on the individual results to arrive at a final ranking.
This is a standard approach and Pyserini adopts the specific implementation in TCT-ColBERT~\cite{Lin_etal_arXiv2020_DenseRanking}, but similar techniques are used elsewhere as well~\cite{karpukhin-etal-2020-dense}.

\subsection{Test Collections}

Beyond the corpus, topics (queries) and relevance judgments (qrels) form indispensable components of IR test collections to support systems-oriented research aimed at producing better ranking models.
Many topics and relevance judgments are freely available for download, but at disparate locations (in various formats)---and often it may not be obvious to a newcomer where to obtain these resources and which exact files to use.

Pyserini tackles this challenge by packaging together these evaluation resources and providing a unified interface for accessing them.
Figure~\ref{fig:queries-qrels-example} shows an example of loading topics via \texttt{get\_topics} (L3) and loading qrels via \texttt{get\_qrels} (L4) for the standard 6980-query subset of the development set of the MS MARCO passage ranking test collection.
We have taken care to name the text descriptors consistently, so the associations between topics and relevance judgments are unambiguous.

\begin{figure}[t]
\small
\begin{tabular}{c}
\begin{lstlisting}
from pyserini.search import get_topics, get_qrels

topics = get_topics('msmarco-passage-dev-subset')
qrels = get_qrels('msmarco-passage-dev-subset')

# Compute the average length of queries:
sum([len(topics[t]['title'].split()) for t in topics])/len(topics)

# Compute the average number of relevance judgments per query:
sum([len(qrels[t]) for t in topics])/len(topics)
\end{lstlisting}
\end{tabular}
\vspace{-0.2cm}
\caption{Simple example of working with queries and qrels from the MS MARCO passage ranking test collection.}
\label{fig:queries-qrels-example}
\vspace{-0.25cm}
\end{figure}

Using Pyserini's provided functions, the topics and qrels are loaded into simple Python data structures and thus easy to manipulate.
A standard TREC topic has different fields (e.g., title, description, narrative), which we model as a Python dictionary.
Similarly, qrels are nested dictionaries:\ query ids mapping to a dictionary of docids to (possibly graded) relevance judgments.
Our choice to use Python data structures means that they can be manipulated using standard constructs such as list comprehensions. 
For example, we can straightforwardly compute the avg.\ length of queries (L7) and the avg.\ number of relevance judgments per query (L10).

\subsection{Batch Retrieval}

Putting everything discussed above together, it is easy in Pyserini to perform an end-to-end batch retrieval run with queries from a standard test collection.
For example, the following command generates a run on the development queries of the MS MARCO passage ranking task (with BM25):

\smallskip
\begin{footnotesize}
\begin{verbatim}
$ python -m pyserini.search --topics msmarco-passage-dev-subset \
   --index msmarco-passage --output run.msmarco-passage.txt \
   --bm25 --msmarco
\end{verbatim}
\end{footnotesize}

\smallskip
\noindent The option \texttt{\mbox{-}-msmarco} specifies the MS MARCO output format; an alternative is the TREC format.
We can evaluate the effectiveness of the run with another simple command:

\smallskip
\begin{footnotesize}
\begin{verbatim}
$ python -m pyserini.eval.msmarco_passage_eval \
   msmarco-passage-dev-subset run.msmarco-passage.txt
#####################
MRR @10: 0.18741227770955546
QueriesRanked: 6980
#####################
\end{verbatim}
\end{footnotesize}

\smallskip
\noindent
Pyserini includes a copy of the official evaluation script and provides a lightweight convenience wrapper around it.
The toolkit manages qrels internally, so the user simply needs to provide the name of the test collection, without having to worry about downloading, storing, and specifying external files.
Otherwise, the usage of the evaluation module is exactly the same as the official evaluation script; in fact, Pyserini simply dispatches to the underlying script after it translates the qrels mapping internally.

The above result corresponds to an Anserini baseline on the MS MARCO passage leaderboard.
This is worth emphasizing and nicely illustrates our goal of making Pyserini easy to use:\ with one simple command, it is possible to replicate a run that serves as a common baseline on a popular leaderboard, providing a springboard to experimenting with different ranking models in a multi-stage architecture.
Similar commands provide replication for batch retrieval with dense representations as well as hybrid retrieval.

\subsection{Working with Custom Collections}

Beyond existing corpora and test collections, a common use case for Pyserini is users who wish to search their own collections.
For bag-of-words sparse retrieval, we have built in Anserini (written in Java) custom parsers and ingestion pipelines for common document formats used in IR research, for example, the TREC SGML format used in many newswire collections and the WARC format for web collections.
However, exposing the right interfaces and hooks to support custom implementations in Python is awkward.
Instead, we have implemented support for a generic and flexible JSON-formatted collection in Anserini (written in Java), and Pyserini's indexer directly accesses the underlying capabilities in An\-serini.
Thus, searching custom collections in Pyserini necessitates first writing a simple script to reformat existing documents into our JSON specification, and then invoking the indexer.
For dense retrieval, support for custom collections is less mature at present, but we provide utility scripts that take an encoder model to convert documents into dense representations, and then build indexes that support querying.

The design of Pyserini makes it easy to use as a standalone module or to integrate as a library in another application.
In the first use case, a researcher can replicate a baseline (first-stage retrieval) run with a simple invocation, take the output run file (which is just plain text) to serve as input for downstream reranking, or as part of  ensembles~\cite{Esteva:2006.09595:2020,Bendersky:2010.00200:2020}.
As an alternative, Pyserini can be used as a library that is tightly integrated into another package; see additional discussions in Section~\ref{section:future}.

\begin{figure}[t]
\small
\begin{tabular}{c}
\begin{lstlisting}
from pyserini.index import IndexReader

# Initialize from a pre-built index:
reader = IndexReader.from_prebuilt_index('robust04')

# Iterate over index terms and fetch term statistics:
import itertools
for term in itertools.islice(reader.terms(), 10):
    print(f'{term.term} (df={term.df}, cf={term.cf})')

# Analyze a term:
term = 'atomic'
analyzed = reader.analyze(term)
print(f'The analyzed form of "{term}" is "{analyzed[0]}"')

# Directly fetch term statistics for a term:
df, cf = reader.get_term_counts(term)
print(f'term "{term}": df={df}, cf={cf}')

# Traverse postings for a term:
postings_list = reader.get_postings_list(term)
for p in postings_list:
    print(f'docid={p.docid}, tf={p.tf}, pos={p.positions}')

# Examples of manipulating document vectors:
tf = reader.get_document_vector('LA071090-0047')
tp = reader.get_term_positions('LA071090-0047')
df = {
    term: (reader.get_term_counts(term, analyzer=None))[0] 
    for term in tf.keys()
}
bm25_vector = {
    term: reader.compute_bm25_term_weight('LA071090-0047',
                                          term,
                                          analyzer=None)
    for term in tf.keys()
}
\end{lstlisting}
\end{tabular}
\vspace{-0.2cm}
\caption{Examples of using Pyserini to access system internals such as term statistics and postings lists.}
\label{fig:internals}
\vspace{-0.4cm}
\end{figure}

\subsection{Access to System Internals}

Beyond simplifying the research lifecycle of working with standard IR test collections, Pyserini provides access to system internals to support use cases that we might not have anticipated.
A number of these features for sparse retrieval are illustrated in Figure~\ref{fig:internals} and available via the \texttt{IndexReader} object, which can be initialized with pre-built indexes in the same way as the searcher classes.\footnote{For these examples, we use the Robust04 index because access to many of the features requires positional indexes and storing document vectors. Due to size considerations, this information is not included in the pre-built MS MARCO indexes.}

In (L7--9), we illustrate how to iterate over all terms in a corpus (i.e., its dictionary) and access each term's document frequency and collection frequency.
Here, we use standard Python tools to select and print out the first 10 terms alphabetically.
In the next example, (L12--14), we show how to ``analyze'' a word (what Lucene calls tokenization, stemming, etc.).
For example, the analyzed form of ``atomic'' is ``atom''.
Since terms in the dictionary (and document vectors, see below) are stored in analyzed form, these methods are necessary to access system internals.
Another way to access collection statistics is shown in (L17--18) by direct lookup.

Pyserini also provides raw access to index structures, both the inverted index as well as the forward index (i.e., to fetch document vectors).
In (L21--23), we show an example of looking up a term's postings list and traversing its postings, printing out term frequency and term position occurrences.
Access to the forward index is shown in (L26--27) based on a docid:\
In the first case, Pyserini returns a dictionary mapping from terms in the document to their term frequencies.
In the second case, Pyserini returns a dictionary mapping from terms to their term positions in the document.
From these methods, we can, for example, look up document frequencies for all terms in a document using a list comprehension in Python (L28--31).
This might be further manipulated to compute tf--idf scores.
Finally, the toolkit provides a convenience method for computing BM25 term weights, using which we can reconstruct the BM25-weighted document vector (L32--37).

At present, access to system internals focuses on manipulating sparse representations.
Dense retrieval capabilities in Pyserini are less mature.
It is not entirely clear what advanced features would be desired by researchers, but we anticipate adding support as the needs and use cases become more clear.

\section{Experimental Results}

Having provided a ``tour'' of Pyserini and some of the toolkit's features, in this section we present experimental results to quantify its effectiveness for first-stage retrieval.
Currently, Pyserini provides support for approximately 30 test collections; here, we focus on two popular leaderboards.

Pyserini provides baselines for two MS MARCO datasets~\cite{MS_MARCO_v3}: the passage ranking task (Table~\ref{result:msmarco-passage}) and the document ranking task (Table~\ref{result:msmarco-doc}).
In both cases, we report the official metric (MRR@10 for passage, MRR@100 for document).
For the development set, we additionally report recall at rank 1000, which is useful in establishing an upper bound on reranking effectiveness.
Note that evaluation results on the test sets are only available via submissions to the leaderboard, and therefore we do not have access to recall figures.
Furthermore, since the organizers discourage submissions that are ``too similar'' (e.g., minor differences in parameter settings) and actively limit the number of submissions to the leaderboard, we follow their guidance and hence do not have test results for all of our experimental conditions.

\begin{table}[t]
\centering\scalebox{0.85}{
\begin{tabular}{llccc}
\toprule
 & & \multicolumn{3}{c}{\textbf{MS MARCO Passage}} \\
 \cmidrule(lr){3-5}
 & & \multicolumn{2}{c}{Development} & Test \\
\multicolumn{2}{l}{\bf Method} & MRR@10 & R@1k & MRR@10 \\
\toprule
\multicolumn{2}{l}{Pyserini:\ sparse} \\
(1a) & Original text & 0.184 & 0.853 & 0.186\\
 & BM25, default ($k_1=0.9, b=0.4$) \\
(1b) & Original text & 0.187 & 0.857 & 0.190 \\
 & BM25, tuned ($k_1=0.82, b=0.68$) \\
(1c) & doc2query--T5 & 0.272 & 0.947 & 0.277 \\
 & BM25, default ($k_1=0.9, b=0.4$) \\
(1d) & doc2query--T5 & 0.282 & 0.951 & - \\
 & BM25, tuned ($k_1=2.18, b=0.86$) \\
\midrule
\multicolumn{2}{l}{Pyserini:\ dense} \\
(2a)  & TCT-ColBERT (brute-force) & 0.335 & 0.964 & - \\
(2b)  & TCT-ColBERT (HNSW) & 0.335 & 0.962 & - \\
\midrule
\multicolumn{2}{l}{Pyserini:\ dense--sparse hybrid} \\
(3a)  & TCT-ColBERT + original text & 0.353 & 0.970 & - \\
(3a)  & TCT-ColBERT + doc2query--T5 & 0.365 & 0.975 & - \\
\midrule
(4a) & BM25 (Microsoft Baseline) & 0.167 & - & 0.165\\
(4b) & ACNE~\cite{Xiong:2007.00808:2020} & 0.330 &  0.959 & - \\
(4c) & DistilBERT$_{\textsc{dot}}$~\cite{Hofstatter:2010.02666:2021} & 0.323 & 0.957 & - \\
\multicolumn{2}{l}{Pyserini:\ multi-stage pipelines} \\
(4d) & monoBERT~\citep{Nogueira_etal_arXiv2019_multistageBERT} & 0.372 & - & 0.365 \\
(4e) & Expando-Mono-DuoT5~\cite{Pradeep_etal_arXiv2021_EMD} & 0.420 & - & 0.408 \\
\bottomrule
\end{tabular}
}
\vspace{0.2cm}
\caption{Results on the MS MARCO passage ranking task.} 
\label{result:msmarco-passage}
\vspace{-0.5cm}
\end{table}

For the passage ranking task, Pyserini supports sparse retrieval, dense retrieval, as well as hybrid dense--sparse retrieval; all results in rows (1) through (3) are replicable with our toolkit.
Row (1a) reports the effectiveness of sparse bag-of-words ranking using BM25 with default parameter settings on the original text; row (1b) shows results after tuning the parameters on a subset of the dev queries via simple grid search to maximize recall at rank 1000.
Parameter tuning makes a small difference in this case.
Pyserini also provides document expansion baselines using our doc2query method~\cite{Nogueira_etal_arXiv2019_doc2query}; the latest model uses T5~\cite{Raffel_etal_JMLR2020} as described in~\citet{Nogueira_Lin_docTTTTTquery}.
Bag-of-words BM25 ranking over the corpus with document expansion is shown in rows (1c) and (1d) for default and tuned parameters.
We see that doc2query yields a large jump in effectiveness, while still using bag-of-words retrieval, since neural inference is applied to generate expansions prior to the indexing phase.
With doc2query, parameter tuning also makes a difference.

For dense retrieval, results using TCT-ColBERT~\cite{Lin_etal_arXiv2020_DenseRanking} are shown in rows (2) using different indexes.
Row (2a) refers to brute-force scans over the document vectors in FAISS~\cite{FAISS}, which provides exact nearest-neighbor search.
Row (2b) refers to approximate nearest-neighbor search using HNSW~\cite{HNSW}; the latter yields a small loss in effectiveness, but enables interactive querying.
We see that retrieval using dense learned representations is much more effective than retrieval using sparse bag-of-words representations, even taking into account document expansion techniques.

Results of hybrid techniques that combine sparse and dense retrieval using weighted interpolation are shown next in Table~\ref{result:msmarco-passage}.
Row (3a) shows the results of combining TCT-ColBERT with BM25 bag-of-words search over the original texts, while row (3b) shows results that combine document expansion using doc2query with the T5 model.
In both cases we used a brute-force approach.
Results show that combining sparse and dense signals is more effective than either alone, and that the hybrid technique continues to benefit from document expansion.

To put these results in context, rows (4) provide a few additional points of comparison.
Row (4a) shows the BM25 baseline provided by the MS MARCO leaderboard organizers, which appears to be less effective than Pyserini's implementation.
Rows (4b) and (4c) refer to two alternative dense-retrieval techniques; these results show that our TCT-ColBERT model performs on par with competing models.
Finally, rows (4d) and (4e) show results from two of our own reranking pipelines built on Pyserini as first-stage retrieval:\ monoBERT, a standard BERT-based reranker~\cite{Nogueira_etal_arXiv2019_multistageBERT}, and our ``Expando-Mono-Duo'' design pattern with T5~\cite{Pradeep_etal_arXiv2021_EMD}.
These illustrate how Pyserini can serve as the foundation for further explorations in neural ranking techniques.

\begin{table}[t]
\centering\scalebox{0.85}{
\begin{tabular}{llccc}
\toprule
 & & \multicolumn{3}{c}{\textbf{MS MARCO Document}} \\
 \cmidrule(lr){3-5}
 & & \multicolumn{2}{c}{Development} & Test \\
\multicolumn{2}{l}{\bf Method} & MRR@100 & R@1k & MRR@100 \\
\toprule
\multicolumn{2}{l}{Pyserini: sparse} \\
(1a) & Original text (doc) & 0.230 & 0.886 & 0.201 \\
 & BM25, default ($k_1=0.9, b=0.4$) \\
(1b) & Original text (doc) & 0.277 & 0.936 & - \\
& BM25, tuned ($k_1=4.46, b=0.82$) \\
(1c) & Original text (passage) & 0.268 & 0.918 & - \\
 & BM25, default ($k_1=0.9, b=0.4$) \\
(1d) & Original text (passage) & 0.275 & 0.931 & 0.246 \\
& BM25, tuned ($k_1=2.16, b=0.61$) \\
(1e) & doc2query--T5 (doc) & 0.327 & 0.955 & 0.291 \\
& BM25, tuned ($k_1=4.68, b=0.87$) \\
(1f) & doc2query--T5 (passage) & 0.321 & 0.953 & 0.290 \\
& BM25, tuned ($k_1=2.56, b=0.59$) \\
\midrule
\multicolumn{2}{l}{Pyserini:\ dense} \\
(2)  & TCT-ColBERT & 0.332 & - & - \\
\midrule
\multicolumn{2}{l}{Pyserini:\ dense--sparse hybrid} \\
(3a)  & TCT-ColBERT + original text  & 0.370 & - & - \\
(3b)  & TCT-ColBERT + doc2query--T5 & 0.378 & - & - \\
\midrule
(4a) & BM25 (Microsoft Baseline) & - & - & 0.192\\
(4b) & ACNE~\cite{Xiong:2007.00808:2020} & 0.384 &  - & 0.342 \\
\multicolumn{2}{l}{Pyserini:\ multi-stage pipelines} \\
(4c) & Expando-Mono-DuoT5~\cite{Pradeep_etal_arXiv2021_EMD} & 0.426 & - & 0.370 \\
\bottomrule
\end{tabular}
}
\vspace{0.2cm}
\caption{Results on the MARCO document ranking task.} 
\label{result:msmarco-doc}
\vspace{-0.5cm}
\end{table}

Results on the MS MARCO document ranking task are shown in Table~\ref{result:msmarco-doc}.
For this task, there are two common configurations, what we call ``per-document'' vs.\ ``per-passage'' indexing.
In the former, each document in the corpus is indexed as a separate document; in the latter, each document is first segmented into multiple passages, and each passage is indexed as a separate ``document''.
Typically, for the ``per-passage'' index, a document ranking is constructed by simply taking the maximum of per-passage scores; the motivation for this design is to reduce the amount of text that computationally expensive downstream rerankers need to process.
Rows (1a)--(1d) show the per-document and per-passage approaches on the original texts, using default parameters and after tuning for recall@100 using grid search.
With default parameters, there appears to be a large effectiveness gap between the per-document and per-passage approaches, but with properly tuned parameters, (1b) vs.\ (1d), we see that they achieve comparable effectiveness.
As with passage retrieval, we can include document expansion with either the per-document or per-passage approaches (the difference is whether we append the expansions to each document or each passage); these results are shown in (1e) and (1f).
Similarly, the differences in effectiveness between the two approaches are quite small.

Dense retrieval using TCT-ColBERT is shown in row (2); this is a new experimental condition that was not reported in~\citet{Lin_etal_arXiv2020_DenseRanking}.
Here, we are simply using the encoder that has been trained on the MS MARCO passage data in a zero-shot manner.
Since these encoders were not designed to process long segments of text, only the per-passage condition makes sense here.
In row (3a), we combine row (2) with the per-passage sparse retrieval results on the original text, and in row (3b), with the per-passage sparse retrieval results using document expansion.
Overall, the findings are consistent with the passage ranking task:\ Dense retrieval is more effective than sparse retrieval (although the improvements for document ranking are smaller, most likely due to  zero-shot application).
Dense and sparse signals are complementary, shown by the effectiveness of the dense--sparse hybrid, which further benefits from document expansion (although the gains from expansion appear to be smaller).

Similar to the passage ranking task, Table~\ref{result:msmarco-doc} provides a few points of comparison.
Row (4a) shows the effectiveness of the BM25 baseline provided by the leaderboard organizers; once again, we see that Pyserini's results are better.
Row (4b) shows ACNE results~\cite{Xiong:2007.00808:2020}, which are more effective than TCT-ColBERT, although the comparison isn't quite fair since our models were not trained on MS MARCO document data.
Finally, Row (4c) shows the results of applying our ``Expando-Mono-Duo'' design pattern with T5~\cite{Pradeep_etal_arXiv2021_EMD} in a zero-shot manner.

In summary, Pyserini ``covers all the bases'' in terms of providing first-stage retrieval for modern research on neural ranking approaches:\ sparse retrieval, dense retrieval, as well as hybrid techniques combining both approaches.
Experimental results on two popular leaderboards show that our toolkit provides a good starting point for further research.

\section{Replicability}
\label{section:replicability}

As replicability is a major consideration in the design and implementation of Pyserini, it is worthwhile to spend some time discussing practices that support this goal.
At a high-level, we can divide replicability into technical and social aspects.
Of the two, we believe the latter are more important, because any technical tool to support replicability will either be ignored or circumvented unless there is a shared commitment to the goal and established social practices to promote it.
Replicability is often in tension with other important desiderata, such as the ability to rapidly iterate, and thus we are constantly struggling to achieve the right balance.

Perhaps the most important principle that our group has internalized is ``to eat our own dog food'', which refers to the colloquialism of using one's own ``product''.
Our group uses Pyserini as the foundation for our own research on transformer-based reranking models, dense learned representations for reranking, and beyond (see more details in Section~\ref{section:future}).
Thus, replicability comes at least partially from our self interest---to ensure that group members can repeat their own experiments and replicate each other's results.
If we can accomplish replicability internally, then external researchers should be able to replicate our results if we ensure that there is nothing peculiar about our computing environment.

Our shared commitment to replicability is operationalized into social processes and is supported by technical infrastructure.
To start, Pyserini as well as the underlying Anserini toolkit adopt standard best practices in open-source software development.
Our code base is available on GitHub, issues are used to describe proposed feature enhancements and bugs, and code changes are mediated via pull requests that are code reviewed by members of our group.

Over the years, our group has worked hard to internalize the culture of writing replication guides for new capabilities, typically paired with our publications; these are all publicly available and stored alongside our code.
These guides include, at a minimum, the sequence of command-line invocations that are necessary to replicate a particular set of experimental results, with accompanying descriptions in prose.
In theory, copying and pasting commands from the guide into a shell should succeed in replication.
In practice, we regularly ``try out'' each other's replication guides to uncover what didn't work and to offer improvements to the documentation.
Many of these guides are associated with a ``replication log'' at the bottom of the guide, which contains a record of individuals who have successfully replicated the results, and the commit id of the code version they used.
With these replication logs, if some functionality breaks, it becomes much easier to debug, by rewinding the code commits back to the previous point where it last ``worked''.

How do we motivate individuals to write these guides and replicate each other's results?
We have two primary tools:\ appealing to reciprocity and providing learning experiences for new group members.
For new students who wish to become involved in our research, conducting replications is an easy way to learn our code base, and hence provides a strong motivation.
In particular, replications are particularly fruitful exercises for undergraduates as their first step in learning about research.
For students who eventually contribute to Pyserini, appeals to reciprocity are effective:\ they are the beneficiaries of previous group members who ``paved the way'' and thus it behooves them to write good documentation to support future students.
Once established, such a culture becomes a self-reinforcing virtuous cycle.

Building on these social processes, replicability in Anserini is further supported by an end-to-end regression framework, that, for each test collection, runs through the following steps:\ builds the index from scratch (i.e., the raw corpus), performs multiple retrieval runs (using different ranking models), evaluates the output (e.g., with \texttt{trec\_eval}), and verifies effectiveness figures against expected results.
Furthermore, the regression framework automatically generates documentation pages from templates, populating results on each successful execution.
All of this happens automatically without requiring any human intervention.
There are currently around 30 such tests, which take approximately two days to run end to end.
The largest of these tests, which occupies most of the time, builds a 12 TB index on all 733 million pages of the ClueWeb12 collection.
Although it is not practical to run these regression tests for each code change, we do try to run them as often as possible, resources permitting.
This has the effect of catching new commits that break existing regressions early so they are easier to debug.
We keep a change log that tracks divergences from expected results (e.g., after a bug fix) or when new regressions are added.

On top of the regression framework in Anserini, further end-to-end regression tests in Pyserini compare its output against An\-se\-rini's output to verify that the Python interface does not introduce any bugs.
These regression tests, for example, test different parameter settings from the command line, ensure that single-threaded and multi-threaded execution yield identical results, that pre-built indexes can be successfully downloaded, etc.

Written guides and automated regression testing lie along a spectrum of replication rigor.
We currently do not have clear-cut criteria as to what features become ``enshrined'' in automated regressions.
However, as features become more critical and foundational in Pyserini or Anserini, we become more motivated to include them in our automated testing framework.

In summary, replicability has become ingrained as a shared norm in our group, operationalized in social processes and facilitated by technical infrastructure.
This has allowed us to balance the demands of replicability with the ability to iterate at a rapid pace.

\section{Future Developments}
\label{section:future}

Anserini has been in development for several years and our group has been working on Pyserini since late 2019.
The most recent major feature added to Pyserini (in 2021) has been dense retrieval capabilities alongside bag-of-words sparse retrieval, and their integration in hybrid sparse--dense techniques.

Despite much activity and continued additions to our toolkit, the broad contours of what Pyserini ``aims to be'' are fairly well defined.
We plan to stay consistent to our goal of providing replicable and easy-to-use techniques that support innovations in neural ranking methods.
Because it is not possible for any single piece of software to do everything, an important part of maintaining focus on our goals is to be clear about what Pyserini {\it isn't going to do}.

While we are planning to add support for more dense retrieval techniques based on learned representations, quite explicitly the {\it training} of these models is outside the scope of Pyserini.
At a high-level, the final ``product'' of any dense retrieval technique comprises an encoder for queries and an encoder for documents (and in some cases, these are the same).
The process of training these encoders can be quite complex, involving, for example, knowledge distillation~\cite{Hofstatter:2010.02666:2021,Lin_etal_arXiv2020_DenseRanking} and complex sampling techniques~\cite{Xiong:2007.00808:2020}.
This is an area of active exploration and it would be premature to try to build a general-purpose toolkit for learning such representations.

For dense retrieval techniques, Pyserini assumes that query/document encoders have already been learned:\ in modern approaches based on pretrained transformers, Huggingface's Transformers library has become the {\it de facto} standard for working with such models, and our toolkit provides tight integration.
From this starting point, Pyserini provides utilities for building indexes that support nearest-neighbor search on these dense representations.
However, it is unlikely that Pyserini will, even in the future, become involved in the training of dense retrieval models.

Another conscious decision we have made in the design of Pyserini is to {\it not} prescribe an architecture for multi-stage ranking and to {\it not} include neural reranking models in the toolkit.
Our primary goal is to provide replicable first-stage retrieval, and we did not want to express an opinion on how multi-stage ranking should be organized.
Instead, our group is working on a separate toolkit, called PyGaggle, that provides implementations for much of our work on multi-stage ranking, including our ``mono'' and ``duo'' designs~\cite{Pradeep_etal_arXiv2021_EMD} as well as ranking with sequence-to-sequence models~\cite{nogueira-etal-2020-document}.
PyGaggle is designed specifically to work with Pyserini, but the latter was meant to be used independently, and we explicitly did not wish to ``hard code'' our own research agenda.
This separation has made it easier for other neural IR toolkits to build on Pyserini, for example, the Caprelous toolkit~\cite{Yates_etal_WSDM2020,Yates_etal_CIKM2020}.

On top of PyGaggle, we have been working on faceted search interfaces to provide a complete end-to-end search application:\ this was initially demonstrated in our Covidex~\cite{ZhangEdwin_etal_SDP2020} search engine for COVID-19 scientific articles.
We have since generalized the application into Cydex, which provides infrastructure for searching the scientific literature, demonstrated in different domains~\cite{Ding_etal_SDP2020}.

Our ultimate goal is to provide reusable libraries for crafting end-to-end information access applications, and we have organized the abstractions in a manner that allows users to pick and choose what they wish to adopt and build on:\ Pyserini to provide first-stage retrieval and basic support, PyGaggle to provide neural re\-ranking models, and Cydex to provide a faceted search interface.

\section{Conclusions}

Our group's efforts to promote and support replicable IR research dates back to 2015~\cite{Lin_etal_ECIR2016,Arguello_etal_SIGIRForum2015}, and the landscape has changed quite a bit since then.
Today, there is much more awareness of the issues surrounding replicability; norms such as the sharing of source code have become more entrenched than before, and we have access to better tools now (e.g., Docker, package mangers, etc.) than we did before.
At the same time, however, today's software ecosystem has become more complex; ranking models have become more sophisticated and modern multi-stage ranking architectures involve more complex components than before.
In this changing environment, the need for stable foundations on which to build remains.
With Pyserini, it has been and will remain our goal to provide easy-to-use tools in support of replicable IR research.

\section*{Acknowledgements}

This research was supported in part by the Canada First Research Excellence Fund, the Natural Sciences and Engineering Research Council (NSERC) of Canada, and the Waterloo--Huawei Joint Innovation Laboratory.

\balance

\bibliographystyle{ACM-Reference-Format}
\bibliography{pyserini}


\begin{thebibliography}{31}


\ifx \showCODEN    \undefined \def \showCODEN     #1{\unskip}     \fi
\ifx \showDOI      \undefined \def \showDOI       #1{#1}\fi
\ifx \showISBNx    \undefined \def \showISBNx     #1{\unskip}     \fi
\ifx \showISBNxiii \undefined \def \showISBNxiii  #1{\unskip}     \fi
\ifx \showISSN     \undefined \def \showISSN      #1{\unskip}     \fi
\ifx \showLCCN     \undefined \def \showLCCN      #1{\unskip}     \fi
\ifx \shownote     \undefined \def \shownote      #1{#1}          \fi
\ifx \showarticletitle \undefined \def \showarticletitle #1{#1}   \fi
\ifx \showURL      \undefined \def \showURL       {\relax}        \fi
\providecommand\bibfield[2]{#2}
\providecommand\bibinfo[2]{#2}
\providecommand\natexlab[1]{#1}
\providecommand\showeprint[2][]{arXiv:#2}

\bibitem[\protect\citeauthoryear{Abadi, Barham, Chen, Chen, Davis, Dean, Devin,
  Ghemawat, Irving, Isard, Kudlur, Levenberg, Monga, Moore, Murray, Steiner,
  Tucker, Vasudevan, Warden, Wicke, Yu, and Zheng}{Abadi et~al\mbox{.}}{2016}]%
        {abadi2016tensorflow}
\bibfield{author}{\bibinfo{person}{Mart{\'\i}n Abadi}, \bibinfo{person}{Paul
  Barham}, \bibinfo{person}{Jianmin Chen}, \bibinfo{person}{Zhifeng Chen},
  \bibinfo{person}{Andy Davis}, \bibinfo{person}{Jeffrey Dean},
  \bibinfo{person}{Matthieu Devin}, \bibinfo{person}{Sanjay Ghemawat},
  \bibinfo{person}{Geoffrey Irving}, \bibinfo{person}{Michael Isard},
  \bibinfo{person}{Manjunath Kudlur}, \bibinfo{person}{Josh Levenberg},
  \bibinfo{person}{Rajat Monga}, \bibinfo{person}{Sherry Moore},
  \bibinfo{person}{Derek~G. Murray}, \bibinfo{person}{Benoit Steiner},
  \bibinfo{person}{Paul Tucker}, \bibinfo{person}{Vijay Vasudevan},
  \bibinfo{person}{Pete Warden}, \bibinfo{person}{Martin Wicke},
  \bibinfo{person}{Yuan Yu}, {and} \bibinfo{person}{Xiaoqiang Zheng}.}
  \bibinfo{year}{2016}\natexlab{}.
\newblock \showarticletitle{{TensorFlow}: A system for large-scale machine
  learning}. In \bibinfo{booktitle}{\emph{Proceedings of the 12th USENIX
  Symposium on Operating Systems Design and Implementation (OSDI '16)}}.
  \bibinfo{pages}{265--283}.
\newblock


\bibitem[\protect\citeauthoryear{{Akkalyoncu Yilmaz}, Clarke, and
  Lin}{{Akkalyoncu Yilmaz} et~al\mbox{.}}{2020}]%
        {Yilmaz_etal_SIGIR2020}
\bibfield{author}{\bibinfo{person}{Zeynep {Akkalyoncu Yilmaz}},
  \bibinfo{person}{Charles L.~A. Clarke}, {and} \bibinfo{person}{Jimmy Lin}.}
  \bibinfo{year}{2020}\natexlab{}.
\newblock \showarticletitle{A Lightweight Environment for Learning Experimental
  {IR} Research Practices}. In \bibinfo{booktitle}{\emph{Proceedings of the
  43rd Annual International ACM SIGIR Conference on Research and Development in
  Information Retrieval (SIGIR 2020)}}. \bibinfo{pages}{2113--2116}.
\newblock


\bibitem[\protect\citeauthoryear{{Akkalyoncu Yilmaz}, Wang, Yang, Zhang, and
  Lin}{{Akkalyoncu Yilmaz} et~al\mbox{.}}{2019}]%
        {Yilmaz_etal_EMNLP2019demo}
\bibfield{author}{\bibinfo{person}{Zeynep {Akkalyoncu Yilmaz}},
  \bibinfo{person}{Shengjin Wang}, \bibinfo{person}{Wei Yang},
  \bibinfo{person}{Haotian Zhang}, {and} \bibinfo{person}{Jimmy Lin}.}
  \bibinfo{year}{2019}\natexlab{}.
\newblock \showarticletitle{Applying {BERT} to Document Retrieval with
  {Birch}}. In \bibinfo{booktitle}{\emph{Proceedings of the 2019 Conference on
  Empirical Methods in Natural Language Processing and the 9th International
  Joint Conference on Natural Language Processing (EMNLP-IJCNLP):\ System
  Demonstrations}}. \bibinfo{address}{Hong Kong, China},
  \bibinfo{pages}{19--24}.
\newblock


\bibitem[\protect\citeauthoryear{Arguello, Crane, Diaz, Lin, and
  Trotman}{Arguello et~al\mbox{.}}{2015}]%
        {Arguello_etal_SIGIRForum2015}
\bibfield{author}{\bibinfo{person}{Jaime Arguello}, \bibinfo{person}{Matt
  Crane}, \bibinfo{person}{Fernando Diaz}, \bibinfo{person}{Jimmy Lin}, {and}
  \bibinfo{person}{Andrew Trotman}.} \bibinfo{year}{2015}\natexlab{}.
\newblock \showarticletitle{Report on the {SIGIR} 2015 Workshop on
  {Reproducibility}, {Inexplicability}, and {Generalizability} of {Results}
  {(RIGOR)}}.
\newblock \bibinfo{journal}{\emph{SIGIR Forum}} \bibinfo{volume}{49},
  \bibinfo{number}{2} (\bibinfo{year}{2015}), \bibinfo{pages}{107--116}.
\newblock


\bibitem[\protect\citeauthoryear{Bajaj, Campos, Craswell, Deng, Gao, Liu,
  Majumder, McNamara, Mitra, Nguyen, Rosenberg, Song, Stoica, Tiwary, and
  Wang}{Bajaj et~al\mbox{.}}{2018}]%
        {MS_MARCO_v3}
\bibfield{author}{\bibinfo{person}{Payal Bajaj}, \bibinfo{person}{Daniel
  Campos}, \bibinfo{person}{Nick Craswell}, \bibinfo{person}{Li Deng},
  \bibinfo{person}{Jianfeng Gao}, \bibinfo{person}{Xiaodong Liu},
  \bibinfo{person}{Rangan Majumder}, \bibinfo{person}{Andrew McNamara},
  \bibinfo{person}{Bhaskar Mitra}, \bibinfo{person}{Tri Nguyen},
  \bibinfo{person}{Mir Rosenberg}, \bibinfo{person}{Xia Song},
  \bibinfo{person}{Alina Stoica}, \bibinfo{person}{Saurabh Tiwary}, {and}
  \bibinfo{person}{Tong Wang}.} \bibinfo{year}{2018}\natexlab{}.
\newblock \showarticletitle{{MS} {MARCO}: {A Human Generated MAchine Reading
  COmprehension Dataset}}.
\newblock \bibinfo{journal}{\emph{arXiv:1611.09268v3}} (\bibinfo{year}{2018}).
\newblock


\bibitem[\protect\citeauthoryear{Bendersky, Zhuang, Ma, Han, Hall, and
  McDonald}{Bendersky et~al\mbox{.}}{2020}]%
        {Bendersky:2010.00200:2020}
\bibfield{author}{\bibinfo{person}{Michael Bendersky}, \bibinfo{person}{Honglei
  Zhuang}, \bibinfo{person}{Ji Ma}, \bibinfo{person}{Shuguang Han},
  \bibinfo{person}{Keith Hall}, {and} \bibinfo{person}{Ryan McDonald}.}
  \bibinfo{year}{2020}\natexlab{}.
\newblock \showarticletitle{{RRF102}: Meeting the TREC-COVID Challenge with a
  100+ Runs Ensemble}.
\newblock \bibinfo{journal}{\emph{arXiv:2010.00200}} (\bibinfo{year}{2020}).
\newblock


\bibitem[\protect\citeauthoryear{Ding, Zhang, and Lin}{Ding
  et~al\mbox{.}}{2020}]%
        {Ding_etal_SDP2020}
\bibfield{author}{\bibinfo{person}{Shane Ding}, \bibinfo{person}{Edwin Zhang},
  {and} \bibinfo{person}{Jimmy Lin}.} \bibinfo{year}{2020}\natexlab{}.
\newblock \showarticletitle{{Cydex}: Neural Search Infrastructure for the
  Scholarly Literature}. In \bibinfo{booktitle}{\emph{Proceedings of the First
  Workshop on Scholarly Document Processing}}. \bibinfo{pages}{168--173}.
\newblock


\bibitem[\protect\citeauthoryear{Esteva, Kale, Paulus, Hashimoto, Yin, Radev,
  and Socher}{Esteva et~al\mbox{.}}{2020}]%
        {Esteva:2006.09595:2020}
\bibfield{author}{\bibinfo{person}{Andre Esteva}, \bibinfo{person}{Anuprit
  Kale}, \bibinfo{person}{Romain Paulus}, \bibinfo{person}{Kazuma Hashimoto},
  \bibinfo{person}{Wenpeng Yin}, \bibinfo{person}{Dragomir Radev}, {and}
  \bibinfo{person}{Richard Socher}.} \bibinfo{year}{2020}\natexlab{}.
\newblock \showarticletitle{{CO-Search}: {COVID-19} Information Retrieval with
  Semantic Search, Question Answering, and Abstractive Summarization}.
\newblock \bibinfo{journal}{\emph{arXiv:2006.09595}} (\bibinfo{year}{2020}).
\newblock


\bibitem[\protect\citeauthoryear{Grand, Muir, Ferenczi, and Lin}{Grand
  et~al\mbox{.}}{2020}]%
        {Grand_etal_ECIR2020}
\bibfield{author}{\bibinfo{person}{Adrien Grand}, \bibinfo{person}{Robert
  Muir}, \bibinfo{person}{Jim Ferenczi}, {and} \bibinfo{person}{Jimmy Lin}.}
  \bibinfo{year}{2020}\natexlab{}.
\newblock \showarticletitle{From {Max\-Score} to {Block-Max} {WAND}: The Story
  of How {Lucene} Significantly Improved Query Evaluation Performance}. In
  \bibinfo{booktitle}{\emph{Proceedings of the 42nd European Conference on
  Information Retrieval, Part II (ECIR 2020)}}. \bibinfo{pages}{20--27}.
\newblock


\bibitem[\protect\citeauthoryear{Hofstätter, Althammer, Schröder, Sertkan,
  and Hanbury}{Hofstätter et~al\mbox{.}}{2021}]%
        {Hofstatter:2010.02666:2021}
\bibfield{author}{\bibinfo{person}{Sebastian Hofstätter},
  \bibinfo{person}{Sophia Althammer}, \bibinfo{person}{Michael Schröder},
  \bibinfo{person}{Mete Sertkan}, {and} \bibinfo{person}{Allan Hanbury}.}
  \bibinfo{year}{2021}\natexlab{}.
\newblock \showarticletitle{Improving Efficient Neural Ranking Models with
  Cross-Architecture Knowledge Distillation}.
\newblock \bibinfo{journal}{\emph{arXiv:2010.02666}} (\bibinfo{year}{2021}).
\newblock


\bibitem[\protect\citeauthoryear{Johnson, Douze, and J{\'e}gou}{Johnson
  et~al\mbox{.}}{2017}]%
        {FAISS}
\bibfield{author}{\bibinfo{person}{Jeff Johnson}, \bibinfo{person}{Matthijs
  Douze}, {and} \bibinfo{person}{Herv{\'e} J{\'e}gou}.}
  \bibinfo{year}{2017}\natexlab{}.
\newblock \showarticletitle{Billion-scale similarity search with GPUs}.
\newblock \bibinfo{journal}{\emph{arXiv:1702.08734}} (\bibinfo{year}{2017}).
\newblock


\bibitem[\protect\citeauthoryear{Karpukhin, Oguz, Min, Lewis, Wu, Edunov, Chen,
  and Yih}{Karpukhin et~al\mbox{.}}{2020}]%
        {karpukhin-etal-2020-dense}
\bibfield{author}{\bibinfo{person}{Vladimir Karpukhin}, \bibinfo{person}{Barlas
  Oguz}, \bibinfo{person}{Sewon Min}, \bibinfo{person}{Patrick Lewis},
  \bibinfo{person}{Ledell Wu}, \bibinfo{person}{Sergey Edunov},
  \bibinfo{person}{Danqi Chen}, {and} \bibinfo{person}{Wen-tau Yih}.}
  \bibinfo{year}{2020}\natexlab{}.
\newblock \showarticletitle{Dense Passage Retrieval for Open-Domain Question
  Answering}. In \bibinfo{booktitle}{\emph{Proceedings of the 2020 Conference
  on Empirical Methods in Natural Language Processing (EMNLP)}}.
  \bibinfo{pages}{6769--6781}.
\newblock


\bibitem[\protect\citeauthoryear{Khattab and Zaharia}{Khattab and
  Zaharia}{2020}]%
        {Khattab_Zaharia_SIGIR2020}
\bibfield{author}{\bibinfo{person}{Omar Khattab} {and} \bibinfo{person}{Matei
  Zaharia}.} \bibinfo{year}{2020}\natexlab{}.
\newblock \showarticletitle{{ColBERT}: Efficient and Effective Passage Search
  via Contextualized Late Interaction over {BERT}}. In
  \bibinfo{booktitle}{\emph{Proceedings of the 43rd International ACM SIGIR
  Conference on Research and Development in Information Retrieval (SIGIR
  2020)}}. \bibinfo{pages}{39--48}.
\newblock


\bibitem[\protect\citeauthoryear{Lin, Crane, Trotman, Callan, Chattopadhyaya,
  Foley, Ingersoll, Macdonald, and Vigna}{Lin et~al\mbox{.}}{2016}]%
        {Lin_etal_ECIR2016}
\bibfield{author}{\bibinfo{person}{Jimmy Lin}, \bibinfo{person}{Matt Crane},
  \bibinfo{person}{Andrew Trotman}, \bibinfo{person}{Jamie Callan},
  \bibinfo{person}{Ishan Chattopadhyaya}, \bibinfo{person}{John Foley},
  \bibinfo{person}{Grant Ingersoll}, \bibinfo{person}{Craig Macdonald}, {and}
  \bibinfo{person}{Sebastiano Vigna}.} \bibinfo{year}{2016}\natexlab{}.
\newblock \showarticletitle{Toward Reproducible Baselines:\ The Open-Source
  {IR} Reproducibility Challenge}. In \bibinfo{booktitle}{\emph{Proceedings of
  the 38th European Conference on Information Retrieval (ECIR 2016)}}.
  \bibinfo{address}{Padua, Italy}, \bibinfo{pages}{408--420}.
\newblock


\bibitem[\protect\citeauthoryear{Lin, Nogueira, and Yates}{Lin
  et~al\mbox{.}}{2020a}]%
        {Lin_etal_arXiv2020_ptr4tr}
\bibfield{author}{\bibinfo{person}{Jimmy Lin}, \bibinfo{person}{Rodrigo
  Nogueira}, {and} \bibinfo{person}{Andrew Yates}.}
  \bibinfo{year}{2020}\natexlab{a}.
\newblock \showarticletitle{Pretrained Transformers for Text Ranking: {BERT}
  and Beyond}.
\newblock \bibinfo{journal}{\emph{arXiv:2010.06467}} (\bibinfo{year}{2020}).
\newblock


\bibitem[\protect\citeauthoryear{Lin, Yang, and Lin}{Lin
  et~al\mbox{.}}{2020b}]%
        {Lin_etal_arXiv2020_DenseRanking}
\bibfield{author}{\bibinfo{person}{Sheng-Chieh Lin},
  \bibinfo{person}{Jheng-Hong Yang}, {and} \bibinfo{person}{Jimmy Lin}.}
  \bibinfo{year}{2020}\natexlab{b}.
\newblock \showarticletitle{Distilling Dense Representations for Ranking using
  Tightly-Coupled Teachers}.
\newblock \bibinfo{journal}{\emph{arXiv:2010.11386}} (\bibinfo{year}{2020}).
\newblock


\bibitem[\protect\citeauthoryear{Malkov and Yashunin}{Malkov and
  Yashunin}{2020}]%
        {HNSW}
\bibfield{author}{\bibinfo{person}{Yu~A. Malkov} {and} \bibinfo{person}{D.~A.
  Yashunin}.} \bibinfo{year}{2020}\natexlab{}.
\newblock \showarticletitle{Efficient and Robust Approximate Nearest Neighbor
  Search Using Hierarchical Navigable Small World Graphs}.
\newblock \bibinfo{journal}{\emph{IEEE Transactions on Pattern Analysis and
  Machine Intelligence}} \bibinfo{volume}{42}, \bibinfo{number}{4}
  (\bibinfo{year}{2020}), \bibinfo{pages}{824--836}.
\newblock


\bibitem[\protect\citeauthoryear{Nogueira, Jiang, Pradeep, and Lin}{Nogueira
  et~al\mbox{.}}{2020}]%
        {nogueira-etal-2020-document}
\bibfield{author}{\bibinfo{person}{Rodrigo Nogueira}, \bibinfo{person}{Zhiying
  Jiang}, \bibinfo{person}{Ronak Pradeep}, {and} \bibinfo{person}{Jimmy Lin}.}
  \bibinfo{year}{2020}\natexlab{}.
\newblock \showarticletitle{Document Ranking with a Pretrained
  Sequence-to-Sequence Model}. In \bibinfo{booktitle}{\emph{Findings of the
  Association for Computational Linguistics: EMNLP 2020}}.
  \bibinfo{pages}{708--718}.
\newblock


\bibitem[\protect\citeauthoryear{Nogueira and Lin}{Nogueira and Lin}{2019}]%
        {Nogueira_Lin_docTTTTTquery}
\bibfield{author}{\bibinfo{person}{Rodrigo Nogueira} {and}
  \bibinfo{person}{Jimmy Lin}.} \bibinfo{year}{2019}\natexlab{}.
\newblock \bibinfo{title}{From doc2query to {docTTTTTquery}}.
\newblock
\newblock


\bibitem[\protect\citeauthoryear{Nogueira, Yang, Cho, and Lin}{Nogueira
  et~al\mbox{.}}{2019a}]%
        {Nogueira_etal_arXiv2019_multistageBERT}
\bibfield{author}{\bibinfo{person}{Rodrigo Nogueira}, \bibinfo{person}{Wei
  Yang}, \bibinfo{person}{Kyunghyun Cho}, {and} \bibinfo{person}{Jimmy Lin}.}
  \bibinfo{year}{2019}\natexlab{a}.
\newblock \showarticletitle{Multi-Stage Document Ranking with {BERT}}.
\newblock \bibinfo{journal}{\emph{arXiv:1910.14424}} (\bibinfo{year}{2019}).
\newblock


\bibitem[\protect\citeauthoryear{Nogueira, Yang, Lin, and Cho}{Nogueira
  et~al\mbox{.}}{2019b}]%
        {Nogueira_etal_arXiv2019_doc2query}
\bibfield{author}{\bibinfo{person}{Rodrigo Nogueira}, \bibinfo{person}{Wei
  Yang}, \bibinfo{person}{Jimmy Lin}, {and} \bibinfo{person}{Kyunghyun Cho}.}
  \bibinfo{year}{2019}\natexlab{b}.
\newblock \showarticletitle{Document Expansion by Query Prediction}.
\newblock \bibinfo{journal}{\emph{arXiv:1904.08375}} (\bibinfo{year}{2019}).
\newblock


\bibitem[\protect\citeauthoryear{Paszke, Gross, Massa, Lerer, Bradbury, Chanan,
  Killeen, Lin, Gimelshein, Antiga, Desmaison, {K\"{o}pf}, Yang, DeVito,
  Raison, Tejani, Chilamkurthy, Steiner, Fang, Bai, and Chintala}{Paszke
  et~al\mbox{.}}{2019}]%
        {paszke2019pytorch}
\bibfield{author}{\bibinfo{person}{Adam Paszke}, \bibinfo{person}{Sam Gross},
  \bibinfo{person}{Francisco Massa}, \bibinfo{person}{Adam Lerer},
  \bibinfo{person}{James Bradbury}, \bibinfo{person}{Gregory Chanan},
  \bibinfo{person}{Trevor Killeen}, \bibinfo{person}{Zeming Lin},
  \bibinfo{person}{Natalia Gimelshein}, \bibinfo{person}{Luca Antiga},
  \bibinfo{person}{Alban Desmaison}, \bibinfo{person}{Andreas {K\"{o}pf}},
  \bibinfo{person}{Edward Yang}, \bibinfo{person}{Zach DeVito},
  \bibinfo{person}{Martin Raison}, \bibinfo{person}{Alykhan Tejani},
  \bibinfo{person}{Sasank Chilamkurthy}, \bibinfo{person}{Benoit Steiner},
  \bibinfo{person}{Lu Fang}, \bibinfo{person}{Junjie Bai}, {and}
  \bibinfo{person}{Soumith Chintala}.} \bibinfo{year}{2019}\natexlab{}.
\newblock \showarticletitle{{PyTorch}: An Imperative Style, High-Performance
  Deep Learning Library}. In \bibinfo{booktitle}{\emph{Advances in Neural
  Information Processing Systems}}. \bibinfo{pages}{8024--8035}.
\newblock


\bibitem[\protect\citeauthoryear{Pradeep, Nogueira, and Lin}{Pradeep
  et~al\mbox{.}}{2021}]%
        {Pradeep_etal_arXiv2021_EMD}
\bibfield{author}{\bibinfo{person}{Ronak Pradeep}, \bibinfo{person}{Rodrigo
  Nogueira}, {and} \bibinfo{person}{Jimmy Lin}.}
  \bibinfo{year}{2021}\natexlab{}.
\newblock \showarticletitle{The Expando-Mono-Duo Design Pattern for Text
  Ranking with Pretrained Sequence-to-Sequence Models}.
\newblock \bibinfo{journal}{\emph{arXiv:2101.05667}} (\bibinfo{year}{2021}).
\newblock


\bibitem[\protect\citeauthoryear{Raffel, Shazeer, Roberts, Lee, Narang, Matena,
  Zhou, Li, and Liu}{Raffel et~al\mbox{.}}{2020}]%
        {Raffel_etal_JMLR2020}
\bibfield{author}{\bibinfo{person}{Colin Raffel}, \bibinfo{person}{Noam
  Shazeer}, \bibinfo{person}{Adam Roberts}, \bibinfo{person}{Katherine Lee},
  \bibinfo{person}{Sharan Narang}, \bibinfo{person}{Michael Matena},
  \bibinfo{person}{Yanqi Zhou}, \bibinfo{person}{Wei Li}, {and}
  \bibinfo{person}{Peter~J. Liu}.} \bibinfo{year}{2020}\natexlab{}.
\newblock \showarticletitle{Exploring the Limits of Transfer Learning with a
  Unified Text-to-Text Transformer}.
\newblock \bibinfo{journal}{\emph{Journal of Machine Learning Research}}
  \bibinfo{volume}{21}, \bibinfo{number}{140} (\bibinfo{year}{2020}),
  \bibinfo{pages}{1--67}.
\newblock


\bibitem[\protect\citeauthoryear{Wolf, Debut, Sanh, Chaumond, Delangue, Moi,
  Cistac, Rault, Louf, Funtowicz, Davison, Shleifer, von Platen, Ma, Jernite,
  Plu, Xu, Le~Scao, Gugger, Drame, Lhoest, and Rush}{Wolf
  et~al\mbox{.}}{2020}]%
        {wolf-etal-2020-transformers}
\bibfield{author}{\bibinfo{person}{Thomas Wolf}, \bibinfo{person}{Lysandre
  Debut}, \bibinfo{person}{Victor Sanh}, \bibinfo{person}{Julien Chaumond},
  \bibinfo{person}{Clement Delangue}, \bibinfo{person}{Anthony Moi},
  \bibinfo{person}{Pierric Cistac}, \bibinfo{person}{Tim Rault},
  \bibinfo{person}{Remi Louf}, \bibinfo{person}{Morgan Funtowicz},
  \bibinfo{person}{Joe Davison}, \bibinfo{person}{Sam Shleifer},
  \bibinfo{person}{Patrick von Platen}, \bibinfo{person}{Clara Ma},
  \bibinfo{person}{Yacine Jernite}, \bibinfo{person}{Julien Plu},
  \bibinfo{person}{Canwen Xu}, \bibinfo{person}{Teven Le~Scao},
  \bibinfo{person}{Sylvain Gugger}, \bibinfo{person}{Mariama Drame},
  \bibinfo{person}{Quentin Lhoest}, {and} \bibinfo{person}{Alexander Rush}.}
  \bibinfo{year}{2020}\natexlab{}.
\newblock \showarticletitle{Transformers: State-of-the-Art Natural Language
  Processing}. In \bibinfo{booktitle}{\emph{Proceedings of the 2020 Conference
  on Empirical Methods in Natural Language Processing: System Demonstrations}}.
  \bibinfo{pages}{38--45}.
\newblock


\bibitem[\protect\citeauthoryear{Xiong, Xiong, Li, Tang, Liu, Bennett, Ahmed,
  and Overwijk}{Xiong et~al\mbox{.}}{2020}]%
        {Xiong:2007.00808:2020}
\bibfield{author}{\bibinfo{person}{Lee Xiong}, \bibinfo{person}{Chenyan Xiong},
  \bibinfo{person}{Ye Li}, \bibinfo{person}{Kwok-Fung Tang},
  \bibinfo{person}{Jialin Liu}, \bibinfo{person}{Paul Bennett},
  \bibinfo{person}{Junaid Ahmed}, {and} \bibinfo{person}{Arnold Overwijk}.}
  \bibinfo{year}{2020}\natexlab{}.
\newblock \showarticletitle{Approximate Nearest Neighbor Negative Contrastive
  Learning for Dense Text Retrieval}.
\newblock \bibinfo{journal}{\emph{arXiv:2007.00808}} (\bibinfo{year}{2020}).
\newblock


\bibitem[\protect\citeauthoryear{Yang, Fang, and Lin}{Yang
  et~al\mbox{.}}{2017}]%
        {Yang_etal_SIGIR2017}
\bibfield{author}{\bibinfo{person}{Peilin Yang}, \bibinfo{person}{Hui Fang},
  {and} \bibinfo{person}{Jimmy Lin}.} \bibinfo{year}{2017}\natexlab{}.
\newblock \showarticletitle{{Anserini}:\ Enabling the Use of {Lucene} for
  Information Retrieval Research}. In \bibinfo{booktitle}{\emph{Proceedings of
  the 40th Annual International ACM SIGIR Conference on Research and
  Development in Information Retrieval (SIGIR 2017)}}. \bibinfo{address}{Tokyo,
  Japan}, \bibinfo{pages}{1253--1256}.
\newblock


\bibitem[\protect\citeauthoryear{Yang, Fang, and Lin}{Yang
  et~al\mbox{.}}{2018}]%
        {Yang_etal_JDIQ2018}
\bibfield{author}{\bibinfo{person}{Peilin Yang}, \bibinfo{person}{Hui Fang},
  {and} \bibinfo{person}{Jimmy Lin}.} \bibinfo{year}{2018}\natexlab{}.
\newblock \showarticletitle{{Anserini}:\ Reproducible Ranking Baselines Using
  {Lucene}}.
\newblock \bibinfo{journal}{\emph{Journal of Data and Information Quality}}
  \bibinfo{volume}{10}, \bibinfo{number}{4} (\bibinfo{year}{2018}),
  \bibinfo{pages}{Article 16}.
\newblock


\bibitem[\protect\citeauthoryear{Yates, Arora, Zhang, Yang, Jose, and
  Lin}{Yates et~al\mbox{.}}{2020a}]%
        {Yates_etal_WSDM2020}
\bibfield{author}{\bibinfo{person}{Andrew Yates}, \bibinfo{person}{Siddhant
  Arora}, \bibinfo{person}{Xinyu Zhang}, \bibinfo{person}{Wei Yang},
  \bibinfo{person}{Kevin~Martin Jose}, {and} \bibinfo{person}{Jimmy Lin}.}
  \bibinfo{year}{2020}\natexlab{a}.
\newblock \showarticletitle{Capreolus: A Toolkit for End-to-End Neural Ad Hoc
  Retrieval}. In \bibinfo{booktitle}{\emph{Proceedings of the 13th ACM
  International Conference on Web Search and Data Mining (WSDM 2020)}}.
  \bibinfo{address}{Houston, Texas}, \bibinfo{pages}{861--864}.
\newblock


\bibitem[\protect\citeauthoryear{Yates, Jose, Zhang, and Lin}{Yates
  et~al\mbox{.}}{2020b}]%
        {Yates_etal_CIKM2020}
\bibfield{author}{\bibinfo{person}{Andrew Yates}, \bibinfo{person}{Kevin~Martin
  Jose}, \bibinfo{person}{Xinyu Zhang}, {and} \bibinfo{person}{Jimmy Lin}.}
  \bibinfo{year}{2020}\natexlab{b}.
\newblock \showarticletitle{Flexible {IR} Pipelines with Capreolus}. In
  \bibinfo{booktitle}{\emph{Proceedings of the 29th International Conference on
  Information and Knowledge Management (CIKM 2020)}}.
  \bibinfo{pages}{3181--3188}.
\newblock


\bibitem[\protect\citeauthoryear{Zhang, Gupta, Tang, Han, Pradeep, Lu, Zhang,
  Nogueira, Cho, Fang, and Lin}{Zhang et~al\mbox{.}}{2020}]%
        {ZhangEdwin_etal_SDP2020}
\bibfield{author}{\bibinfo{person}{Edwin Zhang}, \bibinfo{person}{Nikhil
  Gupta}, \bibinfo{person}{Raphael Tang}, \bibinfo{person}{Xiao Han},
  \bibinfo{person}{Ronak Pradeep}, \bibinfo{person}{Kuang Lu},
  \bibinfo{person}{Yue Zhang}, \bibinfo{person}{Rodrigo Nogueira},
  \bibinfo{person}{Kyunghyun Cho}, \bibinfo{person}{Hui Fang}, {and}
  \bibinfo{person}{Jimmy Lin}.} \bibinfo{year}{2020}\natexlab{}.
\newblock \showarticletitle{{Covidex}: Neural Ranking Models and Keyword Search
  Infrastructure for the {COVID-19} {Open} {Research} {Dataset}}. In
  \bibinfo{booktitle}{\emph{Proceedings of the First Workshop on Scholarly
  Document Processing}}. \bibinfo{pages}{31--41}.
\newblock


\end{thebibliography}

\end{document}